\documentclass[lettersize,journal]{IEEEtran}
\usepackage{amsmath,amsfonts,amssymb}
\usepackage{algorithmic}
\usepackage{algorithm}
\usepackage{array}
\usepackage[caption=false,font=normalsize,labelfont=sf,textfont=sf]{subfig}
\usepackage{textcomp}
\usepackage{stfloats}
\usepackage{url}
\usepackage{verbatim}
\usepackage{graphicx}
\usepackage{cite}
\usepackage{color}
\usepackage{tikz}
\usepackage{hyperref}
\usepackage{geometry}
\usepackage{multirow}
\usepackage{pifont}
\newcommand{\cmark}{\ding{51}}
\newcommand{\xmark}{\ding{55}}
\DeclareMathOperator{\Tr}{Tr}

\newcommand\copyrighttext{%
\footnotesize \textcopyright 2023 IEEE. Personal use of this material is permitted. Permission from IEEE must be obtained for all other uses, in any current or future media, including reprinting/republishing this material for advertising or promotional purposes, creating new collective works, for resale or redistribution to servers or lists, or reuse of any copyrighted component of this work in other works. DOI: \href{https://ieeexplore.ieee.org/document/10352964}{10.1109/LWC.2023.3341254}}
\newcommand\copyrightnotice{%
\begin{tikzpicture}[remember picture,overlay]
\node[anchor=south,yshift=10pt] at (current page.south) {\fbox{\parbox{\dimexpr\textwidth-\fboxsep-\fboxrule\relax}{\copyrighttext}}};
\end{tikzpicture}%
}

\begin{document}

\title{Optimal Window Design for Intercarrier and Intersymbol 
Interference Power Minimization in w-OFDM}

\author{Luiz F. da S. Coelho,~\IEEEmembership{IEEE}, Marcele O. K. Mendonça,~\IEEEmembership{IEEE}, Paulo S. R. Diniz,~\IEEEmembership{IEEE Life Fellow}, Fernando Cruz-Roldán,~\IEEEmembership{Senior Member, IEEE}
        % <-this % stops a space
\thanks{Luiz F. da S. Coelho, Marcele O. K. Mendonça, and Paulo S. R. Diniz are with SMT - Signal, Multimedia, and Telecommunications Laboratory at the Federal University of Rio de Janeiro, 68504, Rio de Janeiro, RJ, 21941-972, Brazil, \{luizfelipe.coelho,~marcele.kuhfuss,~diniz\}@smt.ufrj.br. Fernando Cruz-Rold{\'a}n is with the Department of Teor{\'i}a de la Se{\~n}al y Comunicaciones at the Escuela Polit{\'e}cnica Superior from Universidad de Alcal{\'a}, 28805, Alcal{\'a} de Henares, Spain,~fernando.cruz@uah.es.}% <-this % stops a space
\thanks{This study was financed in part by the Coordenação
de Aperfeiçoamento de Pessoal de Nível Superior - Brasil (CAPES) - Finance Code 001.}
}

\newgeometry{top=.7in,bottom=0.9in,right=0.7in,left=.7in}
\maketitle
\copyrightnotice

\begin{abstract}
Block transceivers are essential for modern communication systems, and they are vastly implemented in many day-to-day devices, especially the orthogonal frequency-division multiplexing (OFDM) technique. The application of a pulse-shaping technique in the OFDM generates the so-called windowed OFDM (w-OFDM). This work proposes an optimization of the window shape, obtaining minimal intercarrier and intersymbol interference power, ICI, and ISI, respectively. A communication system is simulated by implementing the optimized w-OFDM, and its performance is compared with a non-optimized system using a raised-cosine window.
\end{abstract}


\section{Introduction} \label{sec:intro}

\IEEEPARstart{T}{he} conventional cyclic-prefix-based orthogonal frequency division multiplexing (CP-OFDM) has been widely considered in 4G Long-Term Evolution (LTE) due to its ability to deal with multipath fading. CP-OFDM systems combat intersymbol interference (ISI) and intercarrier interference (ICI) caused by the multiple delayed versions of the transmitted signal by using a cyclic prefix (CP) length equal to or greater than the channel order, $\nu$. Although other multicarrier techniques offer better performance in different scenarios, such as orthogonal time-frequency space modulation (OTFS) in high-speed mobile~\cite{wei2021orthogonal} or filter-bank multicarrier (FBMC) for reduced out-of-band radiation (OBR)~\cite{yamada2022system}, OFDM systems are widely considered for 5G~\cite{molisch2023wireless}. This is due to its relatively simple design and low computational cost. By employing time-domain windowing at the transmitter and/or receiver, windowed-OFDM (w-OFDM) can reduce spectrum leakage. OFDM systems with reduced redundancy are attractive alternatives to increase spectrum efficiency further. A particular case is zero-prefix (ZP)-OFDM, which allows a zero-forcing (ZF) solution requiring the minimum amount of redundancy of $\nu/2$ for even $\nu$~\cite{martins2009block,martins2011lti}. Although the ZF solution is not usually discussed for CP-OFDM with insufficient redundancy, the proper window design might enhance the performance.

Several authors have proposed the design of optimized windows for OFDM systems~\cite{Lin05, Lin07,DAl12,Zhu2014,Guv15,Tan2018,hussain2020optimal, Hus22}.  In \cite{Lin05}, the design of optimal windows at the transmitter that minimizes spectral leakage of the transmit signal with the signal-to-noise ratio is presented. The considered OFDM system requires postprocessing at the receiver side. A filterbank transceiver representation is adopted in~\cite{Lin07}, and the receiver window is designed for radio frequency interference reduction. The problem of optimizing the overhead, defined as the roll-off window plus the guard interval, and using a capacity expression as an optimization criterion, is presented in~\cite{DAl12}. An optimal receiver window design over time-varying power-line channels is considered in~\cite{Zhu2014}. The authors propose a maximum signal-to-interference-noise ratio (SINR) criterion using the definition of~\cite{Das2007} over the doubly dispersive channel. A time-asymmetric and per-subcarrier pulse-shaping method to minimize the introduced interference is discussed in~\cite{Guv15}. This approach considers the multipath delay spread as a criterion for pulse-shaping design. In~\cite{Tan2018}, a joint window and filter optimization problem is formulated to minimize out-of-subband emissions. In~\cite{hussain2020optimal,Hus22}, the window design optimization is considered for the weighted overlap-add (WOLA) OFDM with the objective of minimizing the OBR. However, the intersymbol and intercarrier interferences (ISI and ICI, respectively) are neglected.

A criterion for the window design can be either maximizing the SINR or minimizing the interference power. In~\cite{martins2019intersymbol}, the theoretical expressions for the powers of three different types of interference and the SINR for CP-OFDM systems over channels of arbitrary length are obtained. More recently, a unified matrix formulation of CP-OFDM and six different w-OFDM systems (see Table~\ref{system_characteristics}) is derived in~\cite{cruz2020intersymbol}, where theoretical closed-form expressions of interference and noise powers, and the SINR, are also obtained. In addition, different results using conventional RC windows are also shown. Based on the new expressions of~\cite{cruz2020intersymbol}, our main contribution in this letter is the proposal of a novel design technique to obtain windows that minimize total power interference. Although there are some previously proposed methods in which the design is based on reducing the interference or maximizing the SINR, the definitions used by other authors are different from those considered herein. As a novelty, our formulation considers three different types of interference, as well as channels of arbitrary length, where more than one previous symbol may be interfering with the current symbol. In addition, as the new optimization problem is presented using the unified matrix formulation of~\cite{cruz2020intersymbol}, the proposed technique can be applied to design optimal windows for the six different w-OFDM systems shown in Table~\ref{system_characteristics}. This paper also studies the influence of window optimization in these different w-OFDM systems.

\begin{table}[!t]
    \centering
    \caption{Considered w-OFDM systems.}
    \begin{small}
    \begin{tabular}{|c|c|c|c|c|c|} \hline
        {\textbf{System}} & \textbf{Windowing Side} & {\textbf{CP}} & {\textbf{CS}} & \textbf{Reference} \\ \hline \hline
        wtx-OFDM & Tx & \cmark  & \cmark & \cite{DAl12} \\ \hline
        wrx-OFDM & Rx  & \cmark & \cmark & \cite{Mul01} \\ \hline
        WOLA-OFDM & Tx/Rx & \cmark & \cmark & \cite{Zay16}\\ \hline
        CPW-OFDM & Tx/Rx & \cmark & \cmark & \cite{An18}\\ \hline
        CPwtx-OFDM & Tx & \cmark &  \xmark & \cite{Ach11} \\ \hline
        CPwrx-OFDM & Rx & \cmark & \xmark & \cite{Lin07}\\ \hline
    \end{tabular}
    \end{small}
    \label{system_characteristics}
\end{table}

As windowing has a fairly simple computation, the proposed solution was first conceived as a means to mitigate interference in low-power devices, where the window can be stored and used according to the channel state information (CSI). However, devices without strict power restrictions could also benefit from updating the optimal window.

This work is organized as follows. Section~\ref{sec:math} describes the mathematical modeling of the considered w-OFDM systems. In Section~\ref{sec:opt}, the proposed optimization process is presented. Then, the proposed method is evaluated via simulation results in Section~\ref{sec:res}. Finally, Section~\ref{sec:conc} presents some concluding remarks.

\section{Mathematical Modeling} \label{sec:math}

\begin{figure}[!t]
    \centering
    \includegraphics[width=0.85\linewidth]{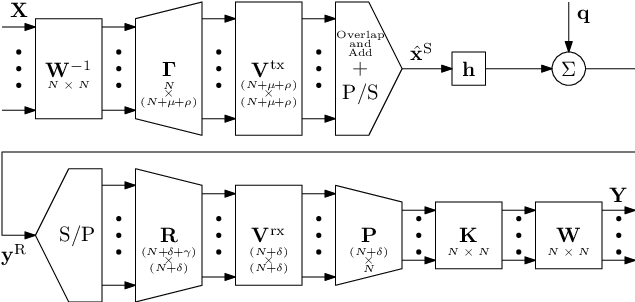}
    \caption{Diagram representation of the studied communication system, using different types of w-OFDM. $\hat{\bf x}^{\textrm{S}}$ is a representation of ${\bf x}^{\textrm{S}}$ after the overlap and add operation.}
    \label{fig:ofdm_diagram}
\end{figure}

In this work, we follow the system model described in~\cite{cruz2020intersymbol}. The OFDM transceiver is depicted in Figure~\ref{fig:ofdm_diagram}, in which the transmitted data vector at the time sample $l \in \mathbb{Z}$ is 
\begin{equation}
     {\bf X}[l] = [X_0[l], \cdots X_k[l] \cdots, X_{N-1}[l]],
\end{equation}
with $N$ being the number of subcarriers. The data samples $X_k[l]$, $k \in \mathcal{N} \triangleq \{0,~1,~\dots,~N-1 \} \subset \mathbb{N}$, belong to a given constellation $\mathcal{C} \subset \mathbb{C}$.

The time-domain transmitted vector can be described as
\begin{align}\label{eq:x_td}
    {\bf x}^\textrm{S}[l] &= {\bf V}^\textrm{tx} \cdot {\bf \Gamma} \cdot {\bf W}^{-1} \cdot {\bf X}[l],
\end{align}
where ${\bf W}^{-1} \in \mathbb{C}^{N \times N}$ is the inverse discrete Fourier transform (IDFT) matrix of size $N$, ${\bf \Gamma} \in \mathbb{N}^{(N+\mu+\rho) \times N}$ introduces the redundancy of $\mu + \rho$, $\mu$ and $\rho$ being the cyclic prefix (CP) and the cyclic suffix (CS) lengths respectively, and ${\bf V}^\textrm{tx} \in \mathbb{R}^{(N+\mu+\rho) \times (N+\mu+\rho)}$ is responsible for the windowing, or pulse-shaping, process at the transmitter.

The windowing matrix is a diagonal matrix 
\begin{align}
    {\bf V}^\textrm{tx} &= \textrm{diag} \left\{ {\bf v}^\textrm{tx} \right\}, \label{eq:window_mat}
\end{align}
where ${\bf v}^\textrm{tx} \in \mathbb{R}^{(N+\mu+\rho) \times 1}$ is a vector containing the time domain window:
\begin{align}
    {\bf v}^{\textrm{tx}} = \left[ {\bf v}^{\textrm{tr}}_{1 \times \beta} \quad {\bf 1}_{1 \times (N+\mu+\rho-2\beta)} \quad {\bf v}^{\textrm{tf}}_{1 \times \beta} \right]^T,
\end{align}
with rising $({\bf v}^{\textrm{tr}})$ and fall $({\bf v}^{\textrm{tf}})$ of $\beta$ samples each and composed by ones $({\bf 1})$ from $\beta$ to $N+\mu+\rho-\beta-1$ and $\textrm{diag}\{ \cdot \}$ is an operator that creates a square matrix containing the operand vector in the diagonal. The overlap and add of $\beta$ samples is modeled in the channel matrix.

Consider that a signal ${\bf x}^\textrm{S} \in \mathbb{C}^{(N+\mu+\rho) \times 1}$, corresponding to a single serialized data vector, is transmitted through the channel and corrupted by additive white Gaussian noise (AWGN). We assume a channel with finite impulse response represented by the vector ${\bf h} = [\rm{h}(0) \quad \rm{h}(1) \quad \cdots \quad \rm{h}(\nu)]$. Furthermore, one can define the number of data vectors that may interfere with the received signal as $M+1$, with
\begin{align}
    M \triangleq \left\lceil \frac{\nu + \beta}{N + \delta + \gamma} \right\rceil,
\end{align}
where $\lceil \cdot \rceil$ is the so-called ceil operator, in which the smallest integer greater than or equal to the operand is the result, $\delta$ is the tail length of the window at the receiving end, and $\gamma$ is the number of samples removed from the beginning of the received signal~\cite{cruz2020intersymbol}.

Hence, assuming perfect synchronization in time and frequency, the received signal ${\bf y}^\textrm{R} \in \mathbb{C}^{(N+\delta+\gamma) \times 1}$ at time $l$ is
\begin{align}\label{eq:y_sum}
    {\bf y}^\textrm{R}[l] &= \sum_{m=0}^{M} {\bf H}^{(m)} \cdot {\bf x}^\textrm{S}[l-m] + {\bf q}[l],
\end{align}
where ${\bf H}^{(m)} \in \mathbb{C}^{(N+\delta+\gamma) \times (N+\mu+\rho)}$ is the channel matrix at the $m$-th past time sample, as defined in~\cite{cruz2020intersymbol}, $M$ is the number of transmitted data that interfere with the received signal and ${\bf q} \in \mathbb{C}^{(N+\delta+\gamma)\times 1}$ is AWGN.

\begin{table}[!t]
    \centering
    \caption{Conditions to Avoid Interference \cite{cruz2020intersymbol}.}
    \begin{small}
    \begin{tabular}{|c|c|} \hline
        {\textbf{System}} & \textbf{Cyclic Prefix Length} \\ \hline \hline
        CP-OFDM &  $\mu \geq \nu $ \\ \hline
        wtx-OFDM &  $\mu \geq \nu + \beta$ \\ \hline
        wrx-OFDM &  $\mu \geq \nu +\frac{\delta}{2}$ \\ \hline
        WOLA-OFDM &  $\mu \geq \nu + \beta + \delta$\\ \hline
        CPW-OFDM &  $\mu \geq \nu + \beta +\frac{\delta}{2}$\\ \hline
        CPwtx-OFDM & $\mu \geq \nu + 2\beta$\\ \hline
        CPwrx-OFDM & $\mu \geq \nu + \delta$\\ \hline
    \end{tabular}
    \end{small}
    \label{no_interference}
\end{table}

If sufficient redundancy is employed (see Table~\ref{no_interference}), only one previously transmitted data will affect the received signal in equation (\ref{eq:y_sum}):
\begin{align}\label{eq:y_m0}
    {\bf y}^{\textrm R}[l] &= {\bf H}^{(0)} \cdot {\bf x}^{\textrm S}[l] + {\bf q}[l].
\end{align}
Conversely, if $\mu$ does not reach the minimum values set out in Table \ref{no_interference} for each system, we have
\begin{align}\label{eq:y_m0_M}
    {\bf y}^\textrm{R}[l] &= {\bf H}^{(0)} \cdot {\bf x}^{\textrm S}[l] +\sum_{m=1}^{M} {\bf H}^{(m)} \cdot {\bf x}^\textrm{S}[l-m] + {\bf q}[l].
\end{align}
${\bf H}^{(0)}$ introduces ICI from the $l$-th symbol, referred to as $\text{ICI}_1$, and ${\bf H}^{(m)},~m>0$, is responsible for the ISI and ICI from the $m$-th previous time sample symbol. The latter is referred to as $\text{ICI}_2$.

At the receiver, the time-domain ${\bf y}^\textrm{R}$ is processed, obtaining in the transform domain
\begin{align}\label{eq:Y_received}
    {\bf Y} [l] &= {\bf W} \cdot {\bf K} \cdot {\bf P} \cdot {\bf V}^{\rm rx} \cdot {\bf R} \cdot {\bf y}^\textrm{R} [l],
\end{align}
where ${\bf R} \in \mathbb{N}^{(N+\delta) \times (N+\delta+\gamma)}$ removes the redundancy by eliminating the first $\gamma$ samples, ${\bf V}^{\rm rx} \in \mathbb{R}^{(N+\delta) \times (N+\delta)}$ represents the windowing process at the receiver, defined as
\begin{align}
    {\bf V}^{\textrm{rx}} = \textrm{diag} \{ {\bf v}^{\textrm{rx}} \},
\end{align}
where ${\bf v}^{\textrm{rx}} \in \mathbb{R}^{(N+\delta) \times 1}$ is a vector containing the time domain window:
\begin{align}
    {\bf v}^{\textrm{rx}} = \begin{bmatrix}
        {\bf v}^{\textrm{rr}}_{1 \times \delta} & {\bf 1}_{1 \times (N-\delta)} & {\bf v}^{\textrm{rf}}_{1 \times \delta}
    \end{bmatrix}^T,
\end{align}
with rising $({\bf v}^{\textrm{rr}})$ and fall $({\bf v}^{\textrm{rf}})$ of $\delta$ real samples each, with values greater than zero and less than one. In addition, ${\bf P} \in \mathbb{N}^{N \times (N+\delta)}$ performs an overlap-and-add operation of the tail samples, which is necessary to recover the transmitted information when the windowing is performed at the receiver; ${\bf K} \in \mathbb{N}^{N \times N}$ is a circular shift necessary for systems with no CS or with shorter CS (WOLA-, CPwtx-, and CPwrx-OFDM); and ${\bf W} \in \mathbb{C}^{N \times N}$ is the discrete Fourier transform (DFT).

Each windowing is conceived differently; while the Tx windowing aims at preparing the block of information for transmission, the Rx windowing selects and shapes the received block envisaging better data recovery. We refer the reader to~\cite{cruz2020intersymbol} for the detailed systems' description and formulation.

\section{Proposed window optimization}
\label{sec:opt}

For a given finite duration channel with arbitrary length $\nu+1$, the received data vector can be expressed in the transform domain as
\begin{align}\label{eq:Y_all}
    {\bf Y}[l] &= \sum^M_{m=0} {\bf A}_m \cdot {\bf X}[l-m] + {\bf G}^{\textrm{noise}} \cdot {\bf q}[l], \\
    {\bf A}_m &= {\bf W} \cdot {\bf K} \cdot {\bf P} \cdot {\bf V}^\textrm{rx} \cdot {\bf R} \cdot {\bf H}^{(m)} \cdot {\bf V}^\textrm{tx} \cdot {\bf \Gamma} \cdot {\bf W}^{-1}, \\
    {\bf G}^{\textrm{noise}} &= {\bf W} \cdot {\bf K} \cdot {\bf P} \cdot {\bf V}^\textrm{rx} \cdot {\bf R}.
\end{align}

Following the same analysis as~\cite{cruz2020intersymbol, martins2019intersymbol}, we assume that the components of the data vector ${\bf X}[l]$ and the noise vector ${\bf q}$ are zero-mean wide-sense stationary uncorrelated processes, independent and identically distributed for all subcarriers, with variances $\sigma^2_X$ and $\sigma^2_q$, respectively. One can define the desired signal component ${\bf Y}_\textrm{des}$ in the received data vector as
\begin{align}\label{eq:y_fd}
    {\bf Y}_\textrm{des}[l] = {\bf A}_0^\textrm{des} \cdot {\bf X}[l],
\end{align}
where ${\bf A}_0^\textrm{des} \in \mathbb{C}^{N \times N}$ is a diagonal matrix with elements $[\mathbf{A}_0^\textrm{des}]_{i,i}=[\mathbf{A}_0]_{i,i}$. The CP works correctly when its length meets the conditions listed in Table \ref{no_interference}. In these cases, there is no interference, ${\bf A}_m=0$, for $m>0$, and the ISI, $\text{ICI}_1$, and $\text{ICI}_2$ effects produced by the wireless channel are null~\cite{cruz2020intersymbol}, and~(\ref{eq:Y_all}) becomes
\begin{align}
    {\bf Y}[l] &= {\bf W} \cdot {\bf B} \cdot {\bf W}^{-1} \cdot {\bf X}[l]+ {\bf G}^{\textrm{noise}} \cdot {\bf q}[l],
\end{align}
where ${\bf B} =  {\bf K} \cdot {\bf P} \cdot {\bf V}^{\rm rx} \cdot {\bf R} \cdot {\bf H}^{(0)} {\bf V}^\textrm{tx} \cdot {\bf \Gamma}$ is right-circulant, and it can be diagonalized by using the  DFT and IDFT matrices. Furthermore, the diagonal matrix ${\bf A}_0 = {\bf W} \cdot {\bf B} \cdot {\bf W}^{-1} = {\bf A}_0^\textrm{des}$. On the contrary, if the conditions of Table \ref{no_interference} are not fulfilled, $Y[l]$ also depends on $X[l-m], m>0$. In addition, ${\bf A}_m \neq 0$ for $m>0$, ${\bf A}_0$ is not diagonal, and three different types of interference occur: $\text{ISI}$, $\text{ICI}_1$ and $\text{ICI}_2$ \cite{cruz2020intersymbol}. In such situations, the system's performance can be enhanced by designing the transmission and the reception windows (${\bf v}^{\rm tx}$ and ${\bf v}^{\rm rx}$) in such a way that the interference power is reduced. From~\cite{cruz2020intersymbol,martins2019intersymbol}, this power is calculated using the ${\bf A}_m$ matrix as follows
\begin{small}
\begin{align}\label{eq:P_tot}
    P_{\textrm{TOT}} =  \sigma_X^2 \cdot \Tr \left[ {\bf A}^{\textrm{ICI}_1} \cdot \left( {{\bf A}^{\textrm{ICI}_1}} \right)^{\rm H} + \sum_{m=1}^M {\bf A}_m \cdot \left( {\bf A}_m \right)^{\rm H} \right],
\end{align}
\end{small}where $\left(\cdot\right)^\textrm{H}$ denotes the Hermitian matrix operation, $\Tr \left[ \cdot \right]$ the trace operator, and ${\bf A}^{\textrm{ICI}_1} = {\bf A}_0 - {\bf A}_0^{\textrm{des}}$. The matrix labeled as ${\bf A}^{\textrm{ICI}_1}$ is responsible for the intercarrier interference of the OFDM block on itself ($\textrm{ICI}_1$), while ${\bf A}_m$ for $m \geq 1$ presents the ISI and $\textrm{ICI}_2$ of other blocks in a given subcarrier $k$. As one may notice, the weights of each window are present in the elements of ${\bf A}_m$; hence, by shaping the windows, we can mitigate the undesired features of ${\bf A}_m$.

We optimize the window once for wtx-, wrx-, CPwtx-, and CPwrx-OFDM systems, where the windowing is performed only at the transmitter or the receiver. Besides, for CPW- and WOLA-OFDM, where we have windowing in both ends, we perform a 3-step process described in the following two cases. In Case A, the first step optimizes the transmitter window considering the receiver with an RC window, the second step optimizes the receiver window considering the optimized transmitter window, and the third step (re)optimizes the transmitter window considering the window obtained in Step 2. In Case B, we perform the same process but starting from the receiver end instead.

Based on \eqref{eq:P_tot}, we propose the minimization of the total power interference taking into account the three types of interference mentioned above: $\text{ICI}_1$ included in ${\bf A}^{\textrm{ICI}_1}$, and $\text{ISI}$ and $\text{ICI}_2$ both of which are contained in ${\bf A}_m$. For this purpose, the objective function is formulated using the squared Frobenius norm $|| \cdot ||_{\textrm{F}}$ of the elements ${\bf A}^{\textrm{ICI}_1}$  and ${\bf A}_m$. Thus, for the transmitter, the optimal window is obtained as
\begin{align}
    &\min_{{\bf v}^\textrm{tx}} \; f({\bf v}^{\textrm{tx}}) = \min_{{\bf v}^\textrm{tx}} \quad \left( ||{\bf A}^{\textrm{ICI}_1}({\bf v}^\textrm{tx})||_\textrm{F}^2  +    \sum_{m=1}^M ||{\bf A}_m({\bf v}^\textrm{tx})||_\textrm{F}^2 \right) \nonumber \\
    &\textrm{subject to:}
    \begin{small}
    \begin{cases}
        {\bf S}^\textrm{H} \cdot {\bf v}^\textrm{tx} = {\bf 1}, \quad \beta+1\leq n \leq N+\mu+\rho- \beta, \\
        0 < {\bf v}^{\textrm{tr}} [n] < 1, \quad 1\leq n \leq \beta, \\
        0 < {\bf v}^{\textrm{tf}}[n] < 1, \quad 1\leq n \leq \beta, \\
        {\bf v}^{\textrm{tr}} [n] = {\bf v}^{\textrm{tf}}[\beta + 1 - n], \quad 1\leq n \leq \beta.
    \end{cases}
    \end{small}\label{eq:cost_tx}
\end{align}
where ${\bf S} \in \mathbb{N}^{(N+\mu+\rho-2\beta) \times (N+\mu+\rho)}$ embraces columns $\beta+1$ through $N+\mu+\rho-\beta$ of the identity matrix ${\bf I}_{N+\mu+\rho}$, and ${\bf 1} \in \mathbb{N}^{(N+\mu+\rho-2\beta) \times 1}$ is an all-ones vector. In \eqref{eq:cost_tx},  the first constraint is included to force the center window samples to be equal to unity. The second and third constraints allow the tail samples to be values between zero and one, while the last constraint forces the window to be symmetric. Similarly, the receiver window optimization problem is defined as
\begin{align}
    &\min_{{\bf v}^\textrm{rx}} \; f({\bf v}^{\textrm{rx}}) =\min_{{\bf v}^\textrm{rx}} \quad \left( ||{\bf A}^{\textrm{ICI}_1}({\bf v}^\textrm{rx})||_\textrm{F}^2  +    \sum_{m=1}^M ||{\bf A}_m({\bf v}^\textrm{rx})||_\textrm{F}^2 \right) \nonumber \\
    &\textrm{subject to:}
    \begin{small}
    \begin{cases}
         {\bf S}^\textrm{H} \cdot {\bf v}^\textrm{rx} = {\bf 1}, \quad \delta+1\leq n \leq N - \delta, \\
         0 < {\bf v}^{\textrm{rr}} [n] < 1, \quad 1\leq n \leq \delta, \\
         0 < {\bf v}^{\textrm{rf}}[n] < 1, \quad 1\leq n \leq \delta, \\
        {\bf v}^{\textrm{rr}} [n] = {\bf v}^{\textrm{rf}}[\delta +1 - n], \quad 1\leq n \leq \delta,  \\
         {\bf v}^{\textrm{rr}} [n] + {\bf v}^{\textrm{rf}}[n] = 1, \quad 1\leq n \leq \delta.       
    \end{cases}
    \end{small}\label{eq:cost_rx}
\end{align}
In \eqref{eq:cost_rx}, there is an additional constraint, which forces the sum of the tail samples used for the reconstruction of a particular sample to be equal to unity. This is a requirement that must be fulfilled for proper system operation. The quadratic formulation of $f({\bf v})$ is derived in the Appendix.

\section{Results}
\label{sec:res}

In our experiments, we consider the six different w-OFDM systems given in Table \ref{system_characteristics} with 256 subcarriers and the parameters shown in Table \ref{parameter_simulat} (see \cite{cruz2020intersymbol}). The input symbols are 16-QAM samples. Each system is tested by simulating a transmission through a set of $250$ independent wireless fading channels mimicking a vehicular communication scenario. The channels are generated using MATLAB's \texttt{stdchan} function, with the channel model \texttt{itur3GVAx} specifying $f_c=2~\textrm{GHz}$ as the carrier frequency, $100~\textrm{km/h}$ as the mobile speed, sample rate of $T_s = 200~\textrm{ns}$, and length $\nu + 1 = 21$. The modeled noise is additive white Gaussian. It is assumed that the channel remains unchanged within the same simulation, and that perfect time and frequency synchronization, and perfect channel estimation, are performed at the receiver.

\begin{figure}[!t]
    \centering
    \includegraphics[width=0.91\linewidth]{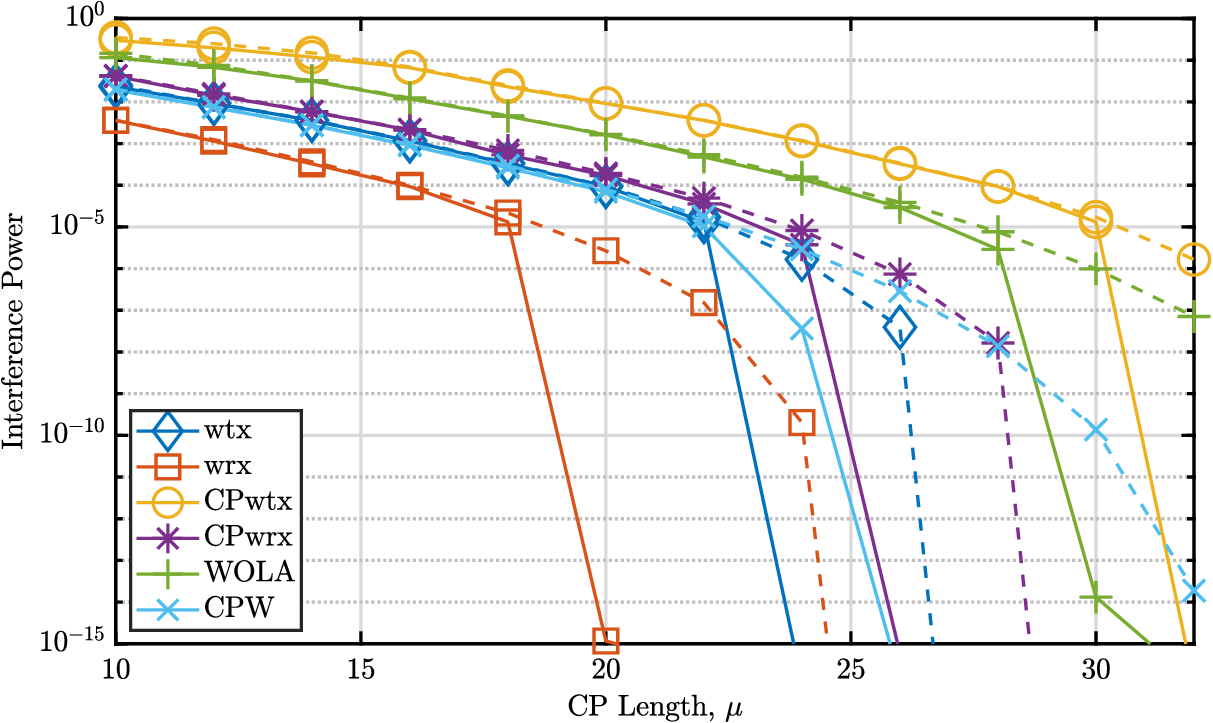}
    \caption{Total interference power for each system. The dashed lines represent systems using RC windows, and the full lines represent the optimized systems.}
    \label{fig:p_isi_ici_final}
\end{figure}

\begin{table}[!t]
\caption{Window parameters used in the experiments ($\mu_{min}$ is the minimum redundancy length to avoid interference for $\nu = 20$).}
    \centering
    \begin{small}
    {\begin{tabular}{|c|c|c|c|c|c|} \hline
        \textbf{System} & $\beta$  & $\delta$ &  $\rho$ & $\gamma$  & $\mu_{min}$ \\ \hline \hline
        \textbf{wtx-OFDM}   & $8$ & $0$ & $\beta$ & $\mu$   & $28$ \\ \hline
        \textbf{wrx-OFDM}  & $0$ & $10$ & $\delta/2$  & $\mu-\delta/2$  & $25$  \\ \hline
        \textbf{WOLA-OFDM}  & $8$ & $10$  & $\beta$   & $\mu - \delta$ & $38$ \\
        \hline
        \textbf{CPW-OFDM}  & $8$ & $10$ & $\beta + \delta/2$ & $\mu-\delta/2$  & 33 \\ \hline
        \textbf{CPwtx-OFDM} & $8$ & $0$ & $0$ & $\mu-\beta$ & 36 \\ \hline
        \textbf{CPwrx-OFDM} & $0$  & $10$  & $0$ & $\mu - \delta$ & $30$\\ \hline
    \end{tabular}}{\label{parameter_simulat}}
    \end{small}
\end{table}

In Figure~\ref{fig:p_isi_ici_final}, the total interference power in equation (\ref{eq:P_tot}) is evaluated for the considered w-OFDM systems when the redundancy length $\mu$ is varied. Here we present the best case scenario for the optimization of WOLA- and CPW-OFDM systems, a 3-step optimization process considering Case B.

Despite having more steps in the optimization, WOLA- and CPW-OFDM systems need longer redundancy than wtx-, CPWrx-, and wrx-OFDM to reach minimum interference. The CP length required to attain minimal interference is $20$ for the optimized wrx-OFDM, while when using the RC window, this is reached with $25$ samples.

\begin{figure}[!t]
    \centering
    \includegraphics[width=.98\linewidth]{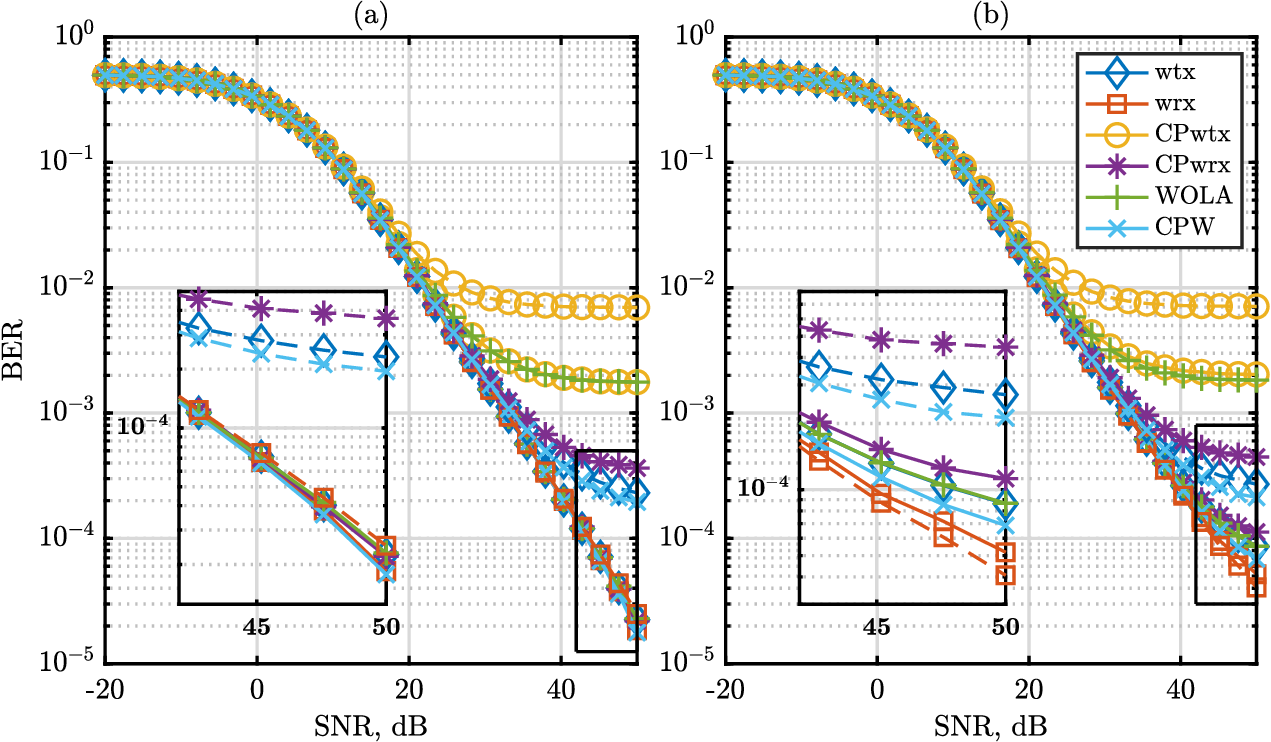}
    \caption{BER curves for w-OFDM systems using $18$ samples in CP. The dashed lines represent systems using RC windows, while the full lines are the optimized systems. (a) has systems without any treatment for OBR, and (b) has an RC channel mask.}
    \label{fig:ber_18cp}
\end{figure}

To indicate that pulse-shaping affects the transmission, the bit-error-rate (BER) curve is presented as a function of the signal-to-noise ratio (SNR) in Figure~\ref{fig:ber_18cp}(a). Since the proposed optimization does not consider the OBR in its cost functions (\ref{eq:cost_tx}) and (\ref{eq:cost_rx}), we simulated systems using channel mask filters as a solution to mitigate the OBR~\cite{pham2016spectrally,zhang2015filtered}, the results are in Figure~\ref{fig:ber_18cp}(b). The mask filter is designed as an RC filter of order $N+\mu+\rho$, and we perform the filtering at the transform domain.

In comparing both scenarios in Figure~\ref{fig:ber_18cp}, the channel mask does not nullify the positive effects of the proposed optimization. In particular, wtx-, WOLA-, CPW-, and CPwrx-OFDM benefit from the optimization process, with a BER reduction for the same values of SNR. However, we observe only a slight improvement in the BER performance of CPwtx-OFDM, since its transmitter window is obtained with a one-step optimization process, and this system needs more redundancy $\left( \mu_{min} =36 \right)$ to fulfill the non-interference condition.

\section{Conclusion}
\label{sec:conc}

This work addresses the issue of designing windows for OFDM systems, including redundancy in various forms. The proposed algorithm is presented under a unified quadratic formulation valid for six different schemes of w-OFDM. The resulting windows reduce the interference in scenarios of insufficient redundancy by shaping the rise and fall tails of the window. In simulations, the behavior of each of these schemes is compared, discussing the effect of optimizing their windows in the BER performance. It is shown that with sub-optimal redundancy, optimized systems can reduce the BER, especially in WOLA- and CPwtx-OFDM systems. Furthermore, it is clear that channel mask filters do not reverse the effects of the optimized window.

\appendix[Cost Functions in Quadratic Form] \label{App:Qform}
The transmitter cost function in \eqref{eq:cost_tx} can be re-written as
\begin{align*}
    f({\bf v}^{\textrm{tx}}) =& \textrm{trace}\left\{ {\bf A}^{\textrm{ICI}_1} \left( {\bf A}^{\textrm{ICI}_1} \right)^H \right\} + \sum_{m=1}^M\textrm{trace}\left\{ {\bf A}_m \left({\bf A}_m\right)^H \right\} \\
    =& \frac{1}{2} \left({\bf v}^{\textrm{tx}} \right)^{T} 2 {\bf Q} {\bf v}^{\textrm{tx}} = \frac{1}{2} \left({\bf v}^{\textrm{tx}} \right)^{T} 2 \left( {\bf Q}_1 + {\bf Q}_2 \right) {\bf v}^{\textrm{tx}}.
\end{align*}

Let us consider the first term of this function, which includes the matrix responsible for $\textrm{ICI}_1$: ${\bf A}^{\textrm{ICI}_1} \triangleq {\bf A}_0 - {\bf A}_0^{\textrm{des}}$. We have
\begin{align}
    {\bf A}_0 = & \underbrace{{\bf W}~{\bf K}~{\bf P}~{\bf V}^{\textrm{rx}}{\bf R}~{\bf H}^{(0)}}_{{\bf B}}~{\bf V}^{\textrm{tx}}~\underbrace{{\bf \Gamma}~{\bf W}^{-1}}_{{\bf C}}, \nonumber \\
    \left\{{\bf A}_0\right\}_{ij} &= \sum_{n=0}^{N+\mu+\rho-1} b_{in} c_{nj} v^{\textrm{tx}}(n), \nonumber
\end{align}
where $b_{in}$ is the element from row $i$ and column $n$ from matrix ${\bf B} \in \mathbb{C}^{N \times (N+\mu+\rho)}$, $c_{nj}$ is the element from the $n$-th row and $j$-th columns in matrix ${\bf C} \in \mathbb{C}^{(N+\mu+\rho) \times N}$, and $v^{\textrm{tx}}(n)$ is the $n$-th element of the ${\bf v}^{\textrm{tx}} \in \mathbb{R}^{N+\mu+\rho}$. Since $\mathbf{A}_0^\textrm{des}=\textrm{diag} \{ \textrm{diag} \{\mathbf{A}_0\} \}$, it turns out that
\begin{align*}
    \left\{ {\bf A}^{\textrm{ICI}_1} \right\}_{ij} &=
    \begin{cases}
        0, \quad & i=j, \\
        \sum_{n=0}^{N+\mu+\rho-1} b_{in} c_{nj} v^{\textrm{tx}}(n), \quad & i \neq j.
    \end{cases}
\end{align*}
\noindent The square Frobenius norm is obtained as
\begin{align}
    \left| \left| {\bf A}^{\textrm{ICI}_1} \right| \right|_F^2 &= \sum_{i=0}^{N-1} \sum_{j=0}^{N-1} \{{\bf A}^{\textrm{ICI}_1}\}_{ij} \{{\bf A}^{\textrm{ICI}_1}\}_{ij}^{\ast}, \quad i \neq j, \nonumber
\end{align}
and the desired sum for the quadratic problem is
\begin{equation*}
    \left({\bf v}^{\textrm{tx}} \right)^{T} {\bf Q}_1 {\bf v}^{\textrm{tx}} = \sum_{n=0}^{N+\mu + \rho - 1} \sum_{m=0}^{N+\mu + \rho - 1} \left\{ {\bf Q}_1 \right\}_{nm} v ^{\textrm{tx}} (n) v ^{\textrm{tx}} (m),
\end{equation*}
where ${\bf Q}_1 \in \mathbb{R}^{(N+\mu+\rho)\times(N+\mu+\rho)}$ with elements
\begin{equation*}
    \left\{ {\bf Q}_1 \right\}_{ij} = \sum_{m=0}^{N-1} \sum_{n=0}^{N-1} b_{mi} c_{in} b_{mj}^{\ast} c_{jn}^{\ast}, \quad m \neq n. \label{eq:q1_ij}
\end{equation*}
Let us consider now the second term of the cost function $f({\bf v}^{\textrm{tx}})$. As the trace operator enables the interchangeability of operation order, considering only the dimensions of the matrices, the elements in $\sum_{m=1}^M || {\bf A}_m ||_F^2$ can be reordered. We define the matrix ${\bf D} = {\bf \Gamma} {\bf K} {\bf P} {\bf V}^{\textrm{rx}} {\bf R} \left( \sum_{m=1}^M {\bf H}^{(m)} \right)$ that concentrates all the operations except the transmitter window. Thus we have
\begin{small}
\begin{align*}
     \sum_{m=1}^M \textrm{trace}\left\{ {\bf A}_m \left( {\bf A}_m \right)^H \right\} &= \textrm{trace} \left\{ {\bf D} {\bf D}^H \left( {\bf V}^{\textrm{tx}} \right) ^2 \right\} \\ 
     &= \sum_{m=0}^{N+\mu + \rho - 1} d_{mm} \left( v ^{\textrm{tx}} (m) \right) ^{2}.
\end{align*}
\end{small}
The elements of $\mathbf{Q}_2=\textrm{diag} \{ \textrm{diag} \{\mathbf{D} \mathbf{D}^H \} \}$ are 
\[\{{\bf Q}_2\}_{i,i} = \{ {\bf DD}^H \}_{i,i}=d_{ii},
\]
where ${\bf Q}_2 \in \mathbb{R}^{(N+\mu+\rho)\times (N+\mu+\rho)}$.

Finally, ${\bf Q} = {\bf Q}_1 + {\bf Q}_2$, where both matrices are real-valued and symmetric and $2{\bf Q}$ (the Hessian matrix) is positive semidefinite implying the convexity of the cost function. \\
The same reasoning applies to obtain ${\bf Q}$ for $f({\bf v}^{\textrm{rx}})$ in \eqref{eq:cost_rx}. In this case, we have ${\bf B} = {\bf W} {\bf K} {\bf P}$, ${\bf C} = {\bf R} {\bf H}^{(0)} {\bf V}^{\textrm{tx}} {\bf \Gamma} {\bf W}^{-1}$, and ${\bf D} = {\bf R} \left( \sum_{m=1}^M {\bf H}^{(m)} \right) {\bf V}^{\textrm{tx}} {\bf \Gamma} {\bf K} {\bf P}$, with ${\bf B} \in \mathbb{C}^{N \times (N+\delta)}$, ${\bf C} \in \mathbb{C}^{(N+\delta) \times N}$ and ${\bf D} \in \mathbb{C}^{(N+\delta)\times (N+\delta)}$, and the resulting Hessian matrix is also positive semidefinite.

\begin{small}

\end{small}

\end{document}